\documentclass[twocolumn,floatfix,aps,prd,showpacs,amsmath,amssymb,nofootinbib,preprintnumbers,superscriptaddress]{revtex4}

\usepackage{graphicx}

\begin{document}

\title{A numerical study of primordial magnetic field amplification
by inflation-produced gravitational waves}

\author{Sachiko Kuroyanagi} \email[]{s-kuro@a.phys.nagoya-u.ac.jp}
\affiliation{Department of Physics, Nagoya University, Chikusa, Nagoya
  464-8602, Japan}

\author{Hiroyuki Tashiro} \affiliation{Institut d'Astrophysique
  Spatiale, Universit\'e Paris-Sud XI and CNR, Orsay, F-91405, France}

\author{Naoshi Sugiyama} \affiliation{Department of Physics, Nagoya
  University, Chikusa, Nagoya 464-8602, Japan} \affiliation{Institute
  for Physics and Mathematics of the Universe, University of Tokyo,
  5-1-5 Kashiwa-no-ha, Kashiwa City, Chiba 277-8582, Japan}

\begin{abstract}
We numerically study the interaction of inflation-produced magnetic
fields with gravitational waves, both of which originate from quantum
fluctuations during inflation.  The resonance between the magnetic field
perturbations and the gravitational waves has been suggested as a
possible mechanism for magnetic field amplification.  However, some
analytical studies suggest that the effect of the inflationary
gravitational waves is too small to provide significant amplification.
Our numerical study shows more clearly how the interaction affects the
magnetic fields and confirms the weakness of the influence of the
gravitational waves.  We present an investigation based on the
magnetohydrodynamic approximation and take into account the differences
of the Alfven speed.
\end{abstract}

\pacs{98.80.Cq, 98.62.En, 04.30.-w}

\maketitle

\section{Introduction}
Recently, various observations have confirmed the presence of magnetic
fields both in individual galaxies and on even larger scales such as in
clusters of galaxies \cite{mag-obs}.  These magnetic fields have a
typical strength of $10^{-6\sim-7}$G.  Such large and strong magnetic
fields are expected to have been amplified to their present levels by
galactic dynamo theory \cite{dynamo-1}.  However this efficiency is a
matter of some debate, because this mechanism requires seed magnetic
fields greater than $\sim 10^{-20}$G \cite{Giovannini} and existing
theories for generating seed fields do not achieve the required strength
\cite{dynamo-2}. To fill the gap between observations and the dynamo
theory, a large number of theoretical models have been suggested to
generate stronger seed fields or to amplify the weak seed fields
generated from existing theory.  The minimum amplitude necessary in
order for dynamo theory to produce the fields observed at the present
epoch is estimated as approximately $10^{-34}$G at a comoving length of
$10$kpc \cite{Davis}.

In this paper, we focus on the amplification of inflation-produced
magnetic fields by inflation-produced gravitational waves; a mechanism
that was originally suggested by Tsagas (2003) \cite{Tsagas1}.
Providing no exotic physics is assumed, inflation generally predicts
quite small magnetic fields of roughly $10^{-53}$G \cite{Turner}.
Various mechanisms to break the conformal invariance of electrodynamics
have been proposed to generate larger magnetic fields, but these are not
widely accepted since they require the inclusion of new physics beyond
the standard model.  Gravitational generation of the massive Z-boson
field has been suggested as a mechanism to naturally break conformal
invariance within the standard model \cite{Davis,Dimopoulos}.  However,
the strength of the generated magnetic field with a reasonable model of
field evolution is estimated to be only $10^{-29}$G on scales of
$100$pc, which is not large enough to explain observations of the
present day universe.  Originally, Ref. \cite{Tsagas1} reported that
such small seed magnetic fields may be amplified to the order of
$10^{15}$ by amplification induced by gravitational waves.  The key idea
of this amplification mechanism is the resonance between gravitational
waves and magnetic field waves as a result of the interaction between
them, which occurs only when these have the same wavenumber.  Since
inflation generates both gravitational waves and magnetic fields at all
scales, this mechanism is considered to be able to work on the
inflationary magnetic fields.  However, this work has since been
revisited by several papers \cite{Betschart,Zunckel,Fenu} which conclude
that it does not provide such significant amplification.

In contrast to the majority of these studies, which have been performed
analytically, we here tackle the problem with a numerical approach.  The
advantage of such a numerical study is that it enables us to accurately
evaluate this effect under more realistic conditions.  We numerically
track the evolutionary history of the universe as it shifts from the
radiation dominated era to the matter dominated era and finally to the
cosmological constant dominated era.  Since the universe is considered
to have high conductivity after it becomes radiation dominated, we make
use of the magnetohydrodynamic (MHD) approximation \cite{Zunckel}.  We
take into account the Alfven velocity, which determines whether or not
resonance amplification occurs in the MHD framework \cite{Marklund}.  In
addition, numerical calculations allow us to follow the time evolution
of the interaction process.  We clearly show the behavior of the
magnetic fields not only in both the long and short wavelength limits,
but also at the point where each mode crosses the horizon, where the
impact of the gravitational waves is the largest.  Because of this
property that the amplification is most efficient when the mode crosses
the horizon, we only investigate the effect on magnetic fields produced
by inflation which is the only mechanism that is able to generate such
fields outside the horizon.

The outline of this paper is as follows.  In Sec. \ref{basics}, we first
introduce the basics of the covariant formalism which is convenient in
describing the interaction between gravitational waves and magnetic
fields.  We then describe equations for the Hubble expansion, magnetic
fields and gravitational waves, which form the base of our
investigation.  In Sec. \ref{calculation}, we first arrange the
equations to be applicable for our numerical calculation, and then
perform the numerical calculation to evaluate the magnitude of the
induced magnetic fields, and show and discuss the results.  We end with
our conclusions in Sec. \ref{conclusion}.

\section{Basic equations}
\label{basics}

\subsection{The covariant formalism}
The interaction between magnetic fields and gravitational waves has been
well studied using the covariant approach. We define the $1+3$
decomposition of space-time by choosing the time direction to be along
the velocity vector of the fundamental observer $u_a$, which satisfies
$u_au^a=-1$. Then the observer's rest frame is described by the
projection tensor $h_{ab}=g_{ab}+u_au_b$. Projecting the covariant
derivative $\nabla_a$ onto the direction of $u_a$ and $h_{ab}$, we
define the time derivative and the spatial derivative as
$\dot{X}_{a\cdots b}\equiv u^c\nabla_c X_{a\cdots b}$ and $D_c
X_{a\cdots b}\equiv{h_c}^d{h_a}^e\cdots{h_b}^f\nabla_d X_{e\cdots f}$,
respectively.  The 3-dimensional curl is defined as ${\rm curl}
X_a\equiv\varepsilon_{abc}D^b X^c$, where $\varepsilon_{abc}$ is the
antisymmetric permutation tensor.

In the covariant formalism, the basic kinematic quantities are defined
by decomposing the covariant derivative of $u_a$,
\begin{equation}
\nabla_bu_a=\sigma_{ab}+\omega_{ab}+\frac{1}{3}\Theta h_{ab}-\dot{u}_au_b,
\end{equation}
where $\sigma_{ab}=D_{( b}u_{a)}$, $\omega_{ab}=D_{[b}u_{a]}$,
$\Theta=\nabla^au_a=D^au_a$ and $\dot{u}_a=u^b\nabla_bu_a$ are
respectively the shear tensor, the vorticity tensor, the volume
expansion scalar, and the 4-acceleration vector. Since we are interested
in cosmological magnetic fields under the influence of gravitational
waves in the expanding universe, we only consider the shear
$\sigma_{ab}$ and the expansion $\Theta$, which correspond to the tensor
perturbation and the Hubble expansion rate, respectively.  Therefore, we
neglect the vorticity $\omega^{ab}$ and acceleration fields $\dot{u}^a$
throughout this paper.

\subsection{Background equation}
For the evolution equation of the background space-time, we assume a
spatially-flat Friedmann Robertson-Walker universe, for which the metric
is given by $ds^2 =-dt^2+a^2(t)\delta_{ij}dx^idx^j$.  In this metric
space, the volume expansion scalar $\Theta$ is related to the scale
factor $a(t)$ as $\Theta=3\dot{a}/a\equiv 3H$, where $H$ is the Hubble
expansion rate and its evolution is simply given by the Friedmann
equation,
\begin{equation}
H=H_0\sqrt{\Omega_r a^{-4}+\Omega_m a^{-3}+\Omega_{\Lambda}}
\label{Hubble},
\end{equation}
where we set $a(t_0)=1$ and the subscript ``0'' denotes the present
time. The density parameters $\Omega_r$, $\Omega_m$, $\Omega_{\Lambda}$
denote radiation, matter, and the cosmological constant, respectively.
We use $\Omega_rh^2=4.15\times10^{-5}$ and the WMAP 5 yr results
$h=0.705$, $\Omega_m=0.274$, and $\Omega_{\Lambda}=0.726$
\cite{WMAP5yr}.

\subsection{Evolution equation for magnetic fields}
The evolution equation for the magnetic field $B_a$ is derived by combining
Maxwell's equations \cite{Tsagas2},
\begin{eqnarray}
&&\dot{E}_{\langle a\rangle}=\sigma_{ab}E^b-\frac{2}{3}\Theta
	    h_{ab}E^b+{\rm curl}B_a-J_{\langle a\rangle},\label{Max1}\\
&&\dot{B}_{\langle a\rangle}=\sigma_{ab}B^b-\frac{2}{3}\Theta
	    h_{ab}B^b-{\rm curl}E_a,\label{Max2}\\
&&D^aE_a=\rho_{\rm e},\\
&&D^aB_a=0,\label{divB}
\end{eqnarray}
where $\rho_{\rm e}=-J_au^a$ is the charge density and $J_{\langle
a\rangle}\equiv h_{ab}J^b$ is the 3-dimensional current.  If we neglect
the effect of the matter components of the universe, $\rho_{\rm e}$ and
$J_{\langle a\rangle}$ can be set to be zero.  This enables us to obtain
the evolution equation for the magnetic fields by only combining the
Maxwell equations \cite{Tsagas1,Tsagas2}, resulting in a description of
the propagation of magnetic field waves in vacuum.

However, the real universe is filled with charged particles, and
therefore we need to take into account the electric current, which is
given by Ohm's law,
\begin{equation}
J_{\langle a\rangle}=\sigma(E_a+\varepsilon_{abc}v^bB^c),
\label{current1}
\end{equation}
where $\sigma$ denotes the electrical conductivity and $v_a$ is the
plasma 3-velocity.  After a large number of charged particles are
produced at the time of reheating following inflation, the conductivity
of the universe is considered to be high enough to assume
$\sigma\rightarrow\infty$, which continues until today even after
recombination of the universe \cite{Turner}.  In this limit, the MHD
approximation is appropriate.  In this paper, we also assume
instantaneous reheating after inflation, allowing us to make use of the
MHD approximation continuously throughout the numerical calculation.

The evolution equation for magnetic fields in the MHD approximation is
derived by using Eq.(\ref{current1}) to replace $E_a$ by $J_a$ in the
Maxwell equations \cite{Marklund}.  In the usual MHD limit, charge
neutrality is assumed so that $\rho_e\approx 0$.  Neglecting the second
time derivative of $B_a$ and taking the limit $\sigma\rightarrow\infty$,
we obtain
\begin{equation}
\dot{B}_{\langle a\rangle}+\frac{2}{3}\Theta
 B_a-\varepsilon_{abc}\varepsilon^{cde}B_eD^bv_d=\sigma_{ab}B^b.
\label{MHDeq}
\end{equation}
The plasma 3-velocity in the third term satisfies the Euler equation,
\begin{equation}
\rho\left[\dot{v}_{\langle
     a\rangle}+v_bD^bv_a+\frac{1}{3}\Theta v_a+\sigma_{ab}v^b\right]=\varepsilon_{abc}J^bB^c,\label{vevo}
\end{equation}
where $\rho$ is the total fluid energy density.  Since the displacement
current $\dot{E}^a$ is negligible compared to the electric current $J_a$
in the MHD limit, Eq.(\ref{Max1}) becomes
\begin{equation}
J_{\langle a\rangle}={\rm curl}B_a\label{current2}.
\end{equation}
Here, we have neglected the expansion term in Eq.(\ref{Max1}), which is
considered to be the same order as $\dot{E}^a$, and also the shear term,
$\sigma_{ab}$, which is much smaller than the expansion $\Theta$.
Linearizing and combining Eqs. (\ref{MHDeq}), (\ref{vevo}) and
(\ref{current2}) yields an evolution equation for the magnetic fields
expressed in terms of only the magnetic field variable $B$,
Eqs. (\ref{Bsevok}), (\ref{Bevok}) and (\ref{hevok}).  The detailed
derivation of this expression is given in Sec. \ref{derivation}.

\subsection{Evolution equation for gravitational waves}
\label{graveq}
In most previous work, the evolution of the gravitational waves is
investigated using the evolution equation of $\sigma_{ab}$
\cite{Tsagas1,Betschart,Zunckel,Fenu}.  However, here we choose to use
the variable $h_{ab}$ instead of $\sigma_{ab}$, since we have thoroughly
investigated the evolution of the gravitational waves using $h_{ij}$ in
our previous work \cite{Kuroyanagi}.  This variable is defined as the
tensor perturbation in a spatially-flat Friedmann Robertson-Walker
universe, $ds^2=-dt^2+a^2(t)(\delta_{ij}+h_{ij})dx^idx^j$, and its
evolution equation is given by the Einstein equation,
\begin{equation}
\ddot{h}_{ij}+3H\dot{h}_{ij}-\frac{1}{a^2}\nabla^2h_{ij}=0.
\label{hevo}
\end{equation}
When considering only the transverse traceless components of
$\sigma_{ab}$, which satisfy $D^a\sigma_{ab}=0$, $h_{ab}$ can be related
to $\sigma_{ab}$ as ${\sigma_a}^b=\dot{h}_a{}^b/2$ \cite{Goode}.  Since
$\sigma_{ab}$ is a first order variable, we therefore also treat
$h_{ab}$ as first order.

Note that we have neglected the anisotropic stress term in
Eq.(\ref{hevo}), which is induced by magnetic fields in general
\cite{Caprini}.  This effectively means we assume that the energy of the
magnetic fields is negligibly small compared to that of the
gravitational waves.  This assumption is valid in the initial phase of
our calculation, since we consider a situation in which the
inflation-produced magnetic fields, which are generally small, are
enhanced by the larger gravitational waves.  However, after the power of
the enhanced magnetic fields becomes comparable to the gravitational
waves, the one-sided energy transfer from the gravitational waves to the
magnetic fields may stop and the magnetic fields start to affect the
evolution of the gravitational waves.  Therefore, strictly speaking we
need to take into account the effect of the anisotropic stress term if
the energy of the magnetic fields grows to exceed that of the
gravitational waves.  However, since our primary aim in the present
study is to ascertain whether the magnetic fields can be amplified to
such a significant level at all, (i.e. to the degree suggested by
Ref. \cite{Tsagas1}), we are justified in ignoring the anisotropic
stress term for the time being.

\section{Numerical calculation}
\label{calculation}
\subsection{Linearization and decomposition of the equations}
\label{derivation}
Here, we rearrange the set of equations, Eqs. (\ref{MHDeq}),
(\ref{vevo}) and (\ref{current2}), to a form suitable for numerical
analysis, referring to the procedure of Ref. \cite{Marklund}.  For the
numerical calculation, we use the linearized and decomposed equations.
First, we linearize the equation by splitting the magnetic fields into a
homogeneous background field $\tilde{B}$, which scales as
$\dot{\tilde{B}}=-2\Theta\tilde{B}/3$, and the perturbed variables.  In
addition, we split the perturbed magnetic fields into parallel ${\cal
B}_a$ and perpendicular ${\cal B}$ components.
\begin{equation}
B_a=\tilde{B}\left[(1+{\cal B})e_a+{\cal B}_a\right],
\label{decB}
\end{equation}
where ${\cal B}\equiv(e^a{B}_a-\tilde{B})/\tilde{B}$ and ${\cal
B}_a\equiv N_{ab}B^b/\tilde{B}$.  The spacelike unit vector $e^a$,
which satisfies $e_ae^a=1$, is parallel to the background magnetic field
$\tilde{B}$, and the projection tensor $N_{ab}=h_{ab}-e_ae_b$ is the
metric of the hyper-surface orthogonal to $e^a$.  In this paper,
$\tilde{B}$ is treated as first order and ${\cal B}$ and ${\cal B}_a$
are second order.  Note that the perturbed variables of the magnetic
fields, ${\cal B}$ and ${\cal B}_a$, are normalized by the magnitude of
the background magnetic field $\tilde{B}$.  This means the linearized
equations are valid as long as these values are much smaller than 1.  We
therefore have to be careful that the amplified fields do not exceed the
background field, ${\cal B}, {\cal B}_a>1$, in which case the
linearization method is no longer applicable.  Here, however, for
similar reasons to those stated in Sec. \ref{graveq}, we apply
linearization as a first step since we are primarily interested in
estimating whether the amplification is significant.\footnote{For the
same reasons, we point out that the linearization method in
Ref. \cite{Tsagas1} is also not applicable if significant amplification
truly occurs.  Furthermore, the effect of the anisotropic stress term
should be taken into account as discussed in Sec. \ref{graveq}. }

Similarly, we decompose the shear perturbation into
three parts as
\begin{equation}
{\sigma}_{ab}=\Sigma_{ab}+2\Sigma_{(a}e_{b)}+\left(e_ae_b-\frac{1}{2}N_{ab}\right)\Sigma,
\label{decsigma}
\end{equation}
where
$\Sigma_{ab}\equiv(N_{(a}\!^cN_{b)}\!^d-N_{ab}N^{cd}/2){\sigma}_{cd}$,
$\Sigma_a\equiv N_{ab}{\sigma}^{bc}e_c$ and
$\Sigma\equiv{\sigma}_{ab}e^ae^b$.  Furthermore we split the spatial
derivative operator into $D_{\|}\equiv e^aD_a=d/dz$ and
$D^a_{\perp}\equiv N^{ab}{\rm D}_b$.

Combining Eqs. (\ref{vevo}) and (\ref{current2}) and neglecting 
second order terms of $v_a$ and $\sigma_{ab}$, we obtain
\begin{equation}
\dot{v}_{\langle a\rangle}+\frac{1}{3}\Theta v_a
=\frac{1}{\rho}(B^bD_bB_a-B^bD_aB_b).
\label{dotv}
\end{equation}
Taking the time derivative of Eq. (\ref{MHDeq}),
\begin{eqnarray}
&&\ddot{B}_a+\frac{2}{3}\dot{\Theta}B_a+\frac{2}{3}\Theta\dot{B}_a+B_bD^b\dot{v}_a-B_aD^b\dot{v}_b\nonumber\\
&&-\frac{1}{3}\Theta B_bD^bv_a+\frac{1}{3}\Theta
 B_aD^bv_b+\dot{B}_bD^bv_a-\dot{B}_aD^bv_b\nonumber\\
&&=\dot{\sigma}_{ab}B^b+\sigma_{ab}\dot{B}^b,
\label{ddotB}
\end{eqnarray}
where we neglect combinations of the spatial curvature ${\cal R}$ and
$v_a$ since they are second order.  We then substitute Eq. (\ref{dotv})
into Eq. (\ref{ddotB}) and conduct the linearization by using
Eqs. (\ref{decB}) and (\ref{decsigma}) and neglecting products of ${\cal
B}$, ${\cal B}_a$ with first order variables, $\Sigma$, $\Sigma_a$
$\Sigma_{ab}$, ${\cal R}$ and $v_a$.  This yields
\begin{eqnarray}
&&(\ddot{\cal B}-\frac{2}{3}\Theta\dot{\cal B})e_a+(\ddot{\cal
 B}_a-\frac{2}{3}\Theta\dot{\cal B}_a)+C_A^2[-D^2{\cal
 B}e_a\nonumber\\
&&-(D_{\|}^2{\cal B}_a-D_{\perp a}D_{\|}{\cal B})]+\frac{4}{3}\Theta(e_bD^bv_a-e_aD^bv_b)\nonumber\\
&&=(\dot{\Sigma}-\frac{2}{3}\Theta\Sigma)e_a+(\dot{\Sigma}_a-\frac{2}{3}\Theta
 \Sigma_a),
\end{eqnarray}
where we define the Alfven velocity as $C_A^2\equiv\tilde{B}^2/\rho$.
Here, we have used $\dot{\tilde{B}}=-2\Theta\tilde{B}/3$ and
$\ddot{\tilde{B}}=-2\dot{\Theta}\tilde{B}/3+4\Theta^2\tilde{B}/9$. 
By linearizing Eq. (\ref{MHDeq}), we obtain the equation to rewrite the
term which includes $v_a$ in terms of the magnetic field variables and
the gravitational wave variables,
\begin{equation}
e_bD^bv_a-e_aD^bv_b=\dot{\cal B}e_a+\dot{\cal B}_a-\Sigma e_a-\Sigma_a.
\end{equation}
Substituting this and the relation $D_{\|}{\cal B}=-D_{\perp a}{\cal
B}^a$, which comes from the constraint of Eq. (\ref{divB}), we obtain the
equations for the parallel component to the direction of the background
field $e^a$,
\begin{equation}
\ddot{\cal B}+\frac{2}{3}\Theta\dot{\cal B}-C_A^2D^2{\cal B}
=\dot{\Sigma}+\frac{2}{3}\Theta\Sigma,
\label{parallel}
\end{equation}
and for the perpendicular component,
\begin{eqnarray}
\ddot{\cal B}_a+\frac{2}{3}\Theta\dot{\cal B}_a-C_A^2(D_{\|}^2{\cal
 B}_a+D_{\perp a}D_{\perp b}{\cal B}^b)&&\nonumber\\
=\dot{\Sigma}_a+\frac{2}{3}\Theta\Sigma_a.&&
\label{perpendicular}
\end{eqnarray}

Next, we decompose the equations by introducing the scalar harmonics
$Q_{(k_\perp)}$ which satisfy
\begin{equation}
D_{\perp}^2Q_{(k_\perp)}=-\frac{k_{\perp}^2}{a^2}Q_{(k_\perp)},
\end{equation}
where $k_\perp$ is the physical wavenumber for the fields perpendicular
to $e^a$.  Taking the divergence and rotation of $Q_{(k_\perp)}$, we
define the vector harmonics of even and odd parity,
\begin{eqnarray}
Q_{(k_\perp)}^a=\left(\frac{k_{\perp}}{a}\right)^{-1}D_{\perp}^aQ_{(k_\perp)},\\
\bar{Q}_{(k_\perp)}^a=\left(\frac{k_{\perp}}{a}\right)^{-1}\epsilon^a\!_bD_{\perp}^bQ_{(k_\perp)},
\end{eqnarray}
where $\epsilon^{ab}=\epsilon_{abc}e^c$.  These harmonics satisfy
$D_{\perp}^2Q_{(k_\perp)}^a=-(k_{\perp}/a)^2Q_{(k_\perp)}^a$ and
$D_{\perp}^2\bar{Q}_{(k_\perp)}^a=-(k_{\perp}/a)^2\bar{Q}_{(k_\perp)}^a$.
We then decompose the magnetic field variables in terms of the scalar
and vector harmonics,
\begin{eqnarray}
{\cal B}&=&\sum_{k_\perp}{\cal B}_{(k_\perp)}^SQ_{(k_\perp)},
\nonumber \\
{\cal B}^a&=&\sum_{k_\perp}\left({\cal B}_{(k_\perp)}^VQ_{(k_\perp)}^a+\bar{\cal B}_{(k_\perp)}^V\bar{Q}_{(k_\perp)}^a\right).
\end{eqnarray}
Similarly, the gravitational wave variables are expressed as 
\begin{eqnarray}
\Sigma&=&\sum_{k_\perp}\Sigma_{(k_\perp)}^SQ_{(k_\perp)},\\
\quad\Sigma^a&=&\sum_{k_\perp}\left(\Sigma_{(k_\perp)}^VQ_{(k_\perp)}^a+\bar{\Sigma}_{(k_\perp)}^V\bar{Q}_{(k_\perp)}^a\right).
\end{eqnarray}
We further decompose the variables by introducing the scalar harmonics
for the parallel component $Q_{(k_\|)}$, which satisfy
$D_{\|}^2Q_{(k_\|)}=-(k_{\|}/a)^2 Q_{(k_\|)}$.  Then ${\cal
B}^S_{(k_\perp)}$ can be rewritten as ${\cal
B}^S_{(k_\perp)}=\sum_{k_\|}{\cal B}_{(k_\perp,k_\|)}^SQ_{(k_\|)}$,
similarly for ${\cal B}_{(k_\perp)}^V$, $\bar{\cal B}_{(k_\perp)}^V$,
${\Sigma}^S_{(k_\perp)}$, ${\Sigma}^V_{(k_\perp)}$,
${\bar\Sigma}^V_{(k_\perp)}$.  Hereafter, we drop subscripts
$(k_\perp,k_\|)$.  By applying the above decompositions,
Eqs. (\ref{parallel}) and (\ref{perpendicular}) result in
\begin{equation}
{\ddot{\cal B}}^S+\frac{2}{3}\Theta{\dot{\cal
 B}}^S+C_A^2\frac{k^2}{a^2}{\cal
 B}^S=\dot\Sigma^S+\frac{2}{3}\Theta\Sigma^S,
\label{Bsevok}
\end{equation}
similarly for ${\cal B}^V_{(k_\perp)}$, and
\begin{equation}
{\ddot{\bar{\cal B}}}\!\,^V+\frac{2}{3}\Theta{\dot{\bar{\cal
 B}}}\!\,^V+C_A^2\frac{k_{\|}^2}{a^2}{\bar{\cal B}}^V=\dot{\bar\Sigma}^V
+\frac{2}{3}\Theta{\bar\Sigma}^V.
\label{Bevok}
\end{equation}
This is the final expression of the equation for the magnetic fields,
which we use in our numerical calculation.

For the calculation of the evolution of the gravitational waves, we follow
the procedure of our previous work \cite{Kuroyanagi}.  We use the
Fourier transformed equation of Eq. (\ref{hevo}),
\begin{equation}
\ddot{h}_{\bf k}^{\lambda}+3H\dot{h}_{\bf k}^{\lambda}+\frac{k^2}{a^2}h_{\bf k}^{\lambda}=0,
\label{hevok}
\end{equation}
where
$h_{ij}(x)=\sum_{\lambda=+,\times}\int d^3k (2\pi)^{-3/2}\epsilon_{ij}^{\lambda}(k)h_{\bf
k}^{\lambda}(k)e^{i{\bf x}\cdot{\bf k}}$.  The polarization tensors
$\epsilon_{ij}^{+,\times}$ satisfy symmetric and transverse-traceless
condition and are normalized as
$\sum_{i,j}^{}\epsilon_{ij}^{\lambda}(\epsilon_{ij}^{\lambda^{\prime}})^*=2\delta^{\lambda\lambda^{\prime}}$.

\subsection{Method}
Using Eqs. (\ref{Hubble}), (\ref{Bevok}) and (\ref{hevok}), we
numerically follow the evolution of both the magnetic fields and the
gravitational waves under the influence of cosmic expansion.  Although
the magnetic field variables are decomposed into several parts, here for
simplicity we first follow the evolution of the variable ${\bar{\cal
B}}^V$.  Strictly speaking all field components should be considered.
However, since their amplitudes are thought to be approximately equal,
one variable is sufficient to demonstrate the main result of this paper
-- that the final amplification is extremely small.  The calculation
begins at the early stage of the radiation dominated universe, from
which the MHD approximation is valid until the present epoch.  The
initial condition of the induced magnetic field ${\bar{\cal B}}^V$ is
set to be zero for all scales.  Note that the variable ${\bar{\cal
B}}^V$ is the value divided by background magnetic field $\tilde{B}$.
This allows us to follow the evolution of the magnetic field without
assuming the strength of the background field.  It may not be reasonable
to assume a homogeneous inflation-produced magnetic field as such fields
are predicted to be a stochastic.  However, as mentioned in
Sec. \ref{graveq}, our primary aim is to ascertain whether the magnetic
fields can be amplified or not as suggested in Ref. \cite{Tsagas1}.
Thus following their assumptions, we evaluate this effect under the
simple situation of a homogeneous background field.  In addition,
although the considered situation is a little different, as the current
term is neglected, the case of a stochastic magnetic field is
investigated in detail analytically in Ref. \cite{Fenu}. These authors
also conclude that significant amplification does not occur.

The initial condition of the gravitational waves is derived from our
numerical calculation presented in Ref. \cite{Kuroyanagi}, which gives
an almost scale-invariant power spectrum.  We define the power spectrum
of the gravitational waves as
\begin{equation}
{\cal P}_{GW}=\frac{k^3}{2\pi^2}\sum_{\lambda=+,\times}|h_{\bf k}^\lambda|^2.
\end{equation}
One may write the power spectrum of the gravitational waves generated
outside the horizon during inflation as
\begin{equation}
{{\cal P}}_{GW,{\rm
 prim}}(k)=\frac{16}{\pi}\left.\left(\frac{H}{m_{Pl}}\right)^2\right|_{k=aH}.
\label{PTprim}
\end{equation}
In our calculation, we assume chaotic inflation, meaning that CMB
normalization gives $H_{\rm inf}/m_{pl}\sim 10^{-5.1}$.  Here, the
amplitude of the decomposed component ${\bar \Sigma}^V$ is assumed to
contain the entire energy given in Eq. (\ref{PTprim}), which means we
use ${\bar \Sigma}^V=\dot{h}^{\lambda}_{\bf k}$ assuming the two polarization
states are equal.  We also assume that it propagates in the direction of
the background field, setting $k_{\rm grav}=k_{\|}$ in
Eqs. (\ref{Bevok}) and (\ref{hevok}).  This assumption is in fact not
strictly true for stochastic inflationary gravitational waves, meaning
that the true amplitude may be a factor of a few smaller. However, this
does not affect the result significantly.

\subsection{Numerical results and discussion}

\begin{figure}
\begin{center}
\includegraphics[width=0.48\textwidth]{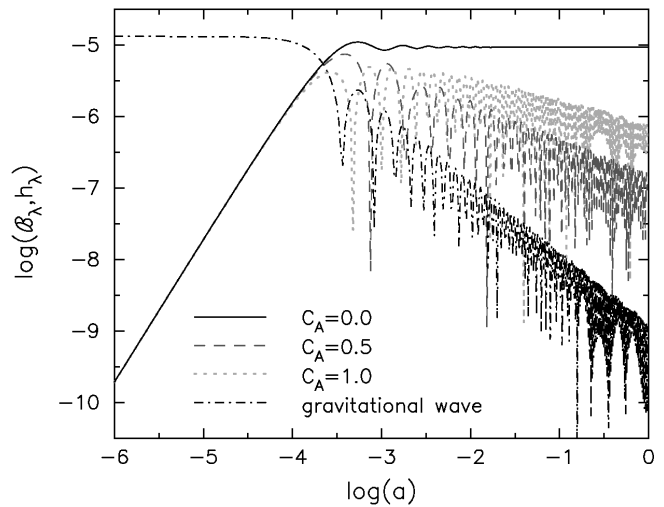} \caption{\label{Bevo}
The evolution of ${\cal B}_{\lambda}\approx\sqrt{{\cal P}_B}$ shown with
the scale factor $a$ as the horizontal axis. Here, we show the mode with
wavenumber $k/a_0H_0=10^2$.  The solid curve shows the case where
$C_A=0.0$, the dashed curve shows $C_A=0.5$, and the dotted curve shows
$C_A=1.0$. Additionally, the dot-dashed line shows the evolution of the
gravitational wave mode $h_{\lambda}=\sqrt{{\cal P}_{GW}}$. }
\end{center}
\end{figure}

\begin{figure}
\begin{center}
\includegraphics[width=0.48\textwidth]{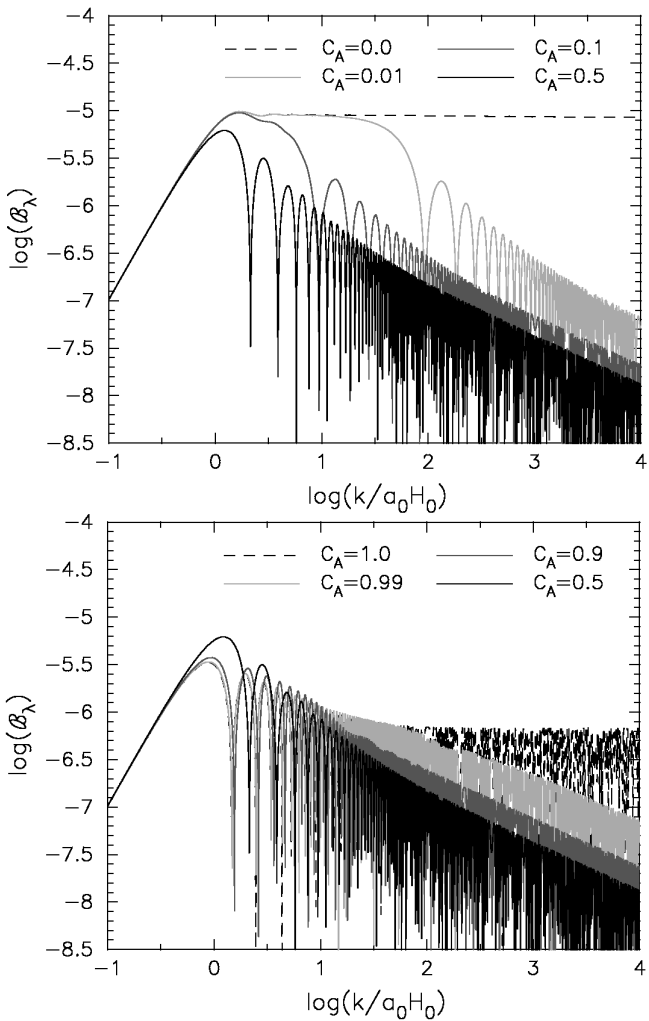} \caption{\label{Bvsk}
The square root of the power spectrum of the induced magnetic fields
${\cal B}_{\lambda}=\sqrt{{\cal P}_B}$ at the present time.  The
horizontal axis is the wavenumber normalized at the present horizon
scale ($\sim 1/a_0H_0$). The upper panel shows the small Alfven speed
case, $C_A=0.0, 0.01, 0.1, 0.5$, and the bottom panel shows the large
Alfven speed case, $C_A=0.5, 0.9, 0.99, 1.0$}
\end{center}
\end{figure}

We first show the time evolution of the square root of the power
spectrum in Fig. \ref{Bevo}, which is defined as
\begin{eqnarray}
{\cal P}_B=\frac{k^3}{2\pi^2}|{\bar{\cal B}}^V|^2.
\end{eqnarray}
This can be interpreted as the power of the induced magnetic field
perturbation with characteristic scale $\lambda=2\pi/k$, ${\cal
B}_{\lambda}\approx\sqrt{{\cal P}_B}$.  In this figure, we show the mode
with wavenumber $k=10^{2}/a_0H_0$ for three cases of different Alfven
speeds, $C_A=0.0, 0.5, 1.0$.  Since the energy density of the magnetic
fields is much smaller than the total energy density of the universe,
the real universe is considered to be close to the case $C_A=0.0$.  We
also plot the evolution of the gravitational wave mode
$h_{\lambda}\approx\sqrt{P_{GW}}$.  As seen in the figure, the largest
amplification occurs at around the point at which a mode enters the
horizon, when the magnetic fields are amplified to the same power as the
gravitational waves.  After the mode enters the horizon, its behavior
depends on the value of the Alfven speed.  For $C_A=0.0$, the induced
magnetic field maintains a constant amplitude, while for $C_A=0.5$ and
$1.0$, the induced field undergoes oscillatory decay, with a decay rate
that depends on the Alfven speed.

According to Eq. (\ref{Bevok}), the behavior described above can be
interpreted as follows.  When the mode is outside the horizon $k\ll aH$,
the evolution of the magnetic fields depends on the amplitude of the
gravitational waves as $\dot{B}_{\bf k}\propto \Sigma_{\bf k}$.  At the
same time, the gravitational waves remain at constant amplitude when
they are outside the horizon, $\Sigma_{\bf k}\sim\dot{h}_{\bf k}\sim
0$. This means the magnetic fields are barely induced at all by those
gravitational waves outside the horizon.  However, when the mode becomes
comparable to the horizon scale ($k\sim aH$), where the gravitational
waves begin to oscillate, $\Sigma_{\bf k}$ becomes larger and induces
magnetic field perturbations.  This is the reason that the magnetic
field perturbation is seen to increase toward the time of horizon
crossing in Fig. \ref{Bevo}.

The evolution inside the horizon $k\gg aH$ depends on the value of the
Alfven speed.  Setting the Alfven velocity to zero is equivalent to
saying there is no Alfven mode.  Thus, the resonant amplification and
the decay of oscillating waves due to cosmic expansion do not occur.
This is the reason why the evolution of the magnetic field fluctuations
in the case $C_A=0.0$ is flat after entering the horizon in
Fig. \ref{Bevo}.  In this case, the amplitude of the induced magnetic
field is determined only by the amplitude of the gravitational waves at
their horizon crossing.  In contrast, if the Alfven mode exists, $C_A
\neq 0.0$, the magnetic field starts to oscillate after entering the
horizon.  In this case, the equation of the magnetic field takes the
form of a forced oscillator sourced by the gravitational wave.  In the
case where the velocities of the magnetic field mode and the
gravitational wave mode are the same, i.e. the case of $C_A=1.0$,
resonant oscillation occurs and pushes up the amplitude of the magnetic
field perturbation.  On the other hand, cosmic expansion reduces the
energy of the oscillating magnetic waves at a rate proportional to
$a^{-1/2}$ and the gravitational waves at a rate proportional to
$a^{-1}$.  Since the gravitational waves decay faster than the magnetic
field perturbations, the amplification effect by resonance oscillation
becomes weaker as the universe evolves.  Therefore, as seen in
Fig. \ref{Bevo}, the magnetic field perturbations decay after they enter
the horizon even when the resonant condition is satisfied.

Figure \ref{Bvsk} shows the square root of the power spectrum at the
present time for several different values of the Alfven speed.  We see
that the induced magnetic fields are the largest in the case of
$C_A=0.0$.  In this case, as mentioned above, there is no resonant
amplification and no decay of the amplitude due to the cosmic expansion
after the horizon crossing, so the amplitude of the induced magnetic
field only depends on the that of the gravitational waves at the horizon
crossing.  Since the primordial power spectrum of the gravitational
waves is scale-invariant, the power spectrum of the induced magnetic
fields also ends up scale-invariant.  On the other hand, in the other
cases where the Alfven speed is not zero, the magnetic fields suffer
damping due to the cosmic expansion after they enter the horizon.  As
seen in the upper panel of Fig. \ref{Bvsk}, the damping arises on modes
smaller than $aH/C_A$.  In the bottom panel, which shows the cases where
the Alfven speed is large enough to bring on the resonance with the
gravitational waves, we see the amplification pushes up the amplitude of
the smaller modes.  The change in the frequency dependence seen around
$k/a_0H_0\sim 10^{1.3}$ is considered to be due to the change of the
Hubble expansion rate at matter-radiation equality, which causes the
change of the frequency dependence of the amplitude of the source term
(gravitational waves).  In any case, the magnitude of the induced
magnetic field is less than $10^{-5}$ times the magnitude of the
original background field at all scales, which leads us to conclude that
the impact of the gravitational waves is negligibly small.

\section{Conclusion}
\label{conclusion}
We have numerically estimated how primordial magnetic fields are
affected by inflation-produced gravitational waves in the framework of
the MHD approximation.  Our calculation has shown that the magnetic
fields gain most energy from the gravitational waves when each mode
enters the horizon, although the impact is not large enough to call it
``amplification''.  We have found that the magnitude of the induced
magnetic field is less than $10^{-5}$ times the magnitude of the
original background field.  This conclusion is consistent with the
analytical estimation in the previous work of
Ref. \cite{Betschart,Zunckel,Fenu}.

Although the induced magnetic fields are quite small in all cases, we
have found that the Alfven speed makes a difference to the evolution of
the magnetic fields inside the horizon.  The Alfven speed is the factor
which determines whether resonance amplification occurs or not, and we
find that it occurs when the Alfven speed is the same as the speed of
light.  However, this situation would not be seen in the real universe,
because such a large Alfven speed corresponds to a situation in which
the magnetic fields dominate the energy density of the universe.
Furthermore, even if we assume this unrealistic situation, our
investigation has found that the amplitude of the induced magnetic
fields does not increase because both the magnetic field waves and the
gravitational waves lose their energy via the cosmic expansion after
they enter the horizon.  On the other hand, if the Alfven speed is
negligibly small, the induced magnetic field does not decay, which
results in a larger amplification rate compared to the case of finite
Alfven speed.

One thing we must note is that our calculation does not take into
account the epoch before reheating, where the conductivity of the
universe is low, and the epoch during reheating, where the Hubble
expansion rate behaves as in a matter dominated universe.  When the
conductivity is low, we cannot apply the MHD equations we have used in
this paper.  However, we do not think the magnetic fields are
significantly affected by the gravitational waves during the low
conductivity epoch, because most of the modes we observe today are
outside the horizon during the early epoch, and the gravitational waves
maintain a constant amplitude.  Although the equation for magnetic
fields in a vacuum is different from that of the MHD limit, the source
term consists of the same components, namely the time derivatives of the
gravitational waves, $\sigma_a^b=\dot{h}_a^b/2$ and $\dot{\sigma}_a^b$
(see Eq. (3) of Ref. \cite{Tsagas1}).  Thus, the gravitational waves
barely induce the magnetic fields outside the horizon, even if the
conductivity is low in the early epoch.  Also, the change in the Hubble
expansion rate does not affect the behavior of the constant
gravitational waves when they are outside the horizon.  Therefore, the
early epoch before and during reheating is considered to not change our
results.

\section*{Acknowledgments}
The authors are grateful to Kiyotomo Ichiki, Keitaro Takahashi and
Christos G. Tsagas for helpful discussion, and to Joanne Dawson for
careful correction of the manuscript.  S.K. would like to thank Koji
Tomisaka and Takahiro Kudo for helpful comment in the early stage of
this work.  This research is supported by Grant-in-Aid for Nagoya
University Global COE Program, "Quest for Fundamental Principles in the
Universe: from Particles to the Solar System and the Cosmos", and
Grant-in-Aid for Scientific Research on Priority Areas No. 467 "Probing
the Dark Energy through an Extremely Wide and Deep Survey with Subaru
Telescope", from the Ministry of Education, Culture, Sports, Science and
Technology of Japan.



\begin{thebibliography}{99}
\bibitem{mag-obs}
A. M. Wolfe, K. M. Lanzetta, and A. L. Oren, Astrophys.
J. {\bf 388}, 17 (1992);
T. E. Clarke, P. P. Kronberg, and H. Boehringer, Astrophys.
J. {\bf 547}, L111 (2001);
L. M. Widrow, Rev. Mod. Phys. {\bf 74}, 775 (2002);
Y. Xu, P. P. Kronberg, S. Habib, and Q. W. Dufton,
Astrophys. J. {\bf 637}, 19 (2006);
P. P. Kronberg et al. Astrophys. J. {\bf 676}, 7079 (2008).

\bibitem{dynamo-1}
A. Brandenburg and K. Subramanian, Phys. Rep. {\bf 417}, 1 (2005).

\bibitem{Giovannini} M. Giovannini, Int. J. Mod. Phys. D {\bf 13}, 391 (2004).

\bibitem{dynamo-2}
R. Kulsrud et al., Phys. Rep. {\bf 283}, 213 (1997).

\bibitem{Davis} A. C. Davis, K. Dimopoulos, T. Prokopec and O. Trnkvist,
	Phys. Lett. B {\bf 501}, 165 (2001).

\bibitem{Tsagas1} C. G. Tsagas, P. K. S. Dunsby and M. Marklund,
	Phys. Lett. B {\bf 561}, 17 (2003); C. G. Tsagas, Phys. Rev. D {\bf
	72}, 123509 (2005); Phys. Rev. D {\bf 75}, 087901 (2007).

\bibitem{Turner} M. S. Turner and L. M. Widrow, Phys. Rev. D {\bf 37},
	2743 (1988).

\bibitem{Dimopoulos} K. Dimopoulos,
	T. Prokopec, O. Tornkvist and A-C. Davis, Phys. Rev. D {\bf 65},
	063505 (2002).

\bibitem{Betschart} G. Betschart, C. Zunckel, P. K. S. Dunsby and
	M. Marklund, Phys. Rev. D {\bf 72}, 123514 (2005); Phys. Rev. D
	{\bf 75}, 087902 (2007).

\bibitem{Zunckel} C. Zunckel, G. Betschart, P. K. S. Dunsby and
	M. Marklund, Phys. Rev. D {\bf 73}, 103509 (2006).

\bibitem{Fenu} E. Fenu and Ruth Durrer, Phys. Rev. D {\bf 79}, 024021
	(2009).

\bibitem{Marklund} M. Marklund and C. Clarkson,
	Mon. Not. Roy. Astron. Soc. {\bf 358}, 892 (2005).

\bibitem{WMAP5yr} E. Komatsu, {\it et al}., Astrophys. J. Suppl. {\bf 180},
	330 (2009).

\bibitem{Tsagas2} C. G. Tsagas, Classical and Quantum Gravity {\bf 22},
	393 (2005).

\bibitem{Kuroyanagi} S. Kuroyanagi, T. Chiba and N. Sugiyama,
	Phys. Rev. D {\bf 79}, 103501 (2009).

\bibitem{Goode} S. W. Goode, Phys. Rev. D {\bf 39}, 2882 (1989).

\bibitem{Caprini} C. Caprini and R. Durrer, Phys. Rev. D {\bf 65},
	023517 (2001); Phys. Rev. D {\bf 74}, 063521 (2006); C. Caprini,
	R. Durrer and G. Servant, arXiv:0909.0622 [astro-ph.CO].

\end{thebibliography}
\end{document}